%% file: jinguogong.tex
\documentclass[11pt]{article}
\usepackage{epsfig,amsmath,amssymb,euscript}
\usepackage{morefloats}
\usepackage{tikz}
\usetikzlibrary{shapes,arrows}
\usetikzlibrary{positioning}

\renewcommand{\baselinestretch}{1.4}
\makeatletter

\input{QYalias}

\newcommand{\Date}[1]{\def\@Date{#1}}
\def\today{\number\day~\ifcase\month\or
 January\or February\or March\or April\or May\or June\or
 July\or August\or September\or October\or November\or December\fi~\number\year}

\def\MARE{\rm MARE}

\begin{document}

\title{\bf Estimation of Extreme Quantiles for Functions of Dependent
Random Variables}

{\small
\author{
Jinguo Gong\\
School of Statistics\\
Southwestern University of Finance\\
 and Economics,
Chengdu, % 611130,
China\\
{\footnotesize jinguogong@swufe.edu.cn}
\and
Yadong Li\\
QA Exposure Analytics\\
Barclays Bank\\
% 5 The North Colonnade, Canary Wharf\\
%London E14 4BB, UK\\
New York, USA\\
{\footnotesize
yadong.li@barclays.com}
\and
Liang Peng\\
School of Mathematics\\
Georgia Institute of Technology\\
% 686 Cherry Street\\
Atlanta,
% GA 30332-0160,
USA\\
{\footnotesize
peng@math.gatech.edu}
\and
Qiwei Yao\\
Department of Statistics\\
London School of Economics\\
% Houghton Street\\
London, UK\\
{\footnotesize
q.yao@lse.ac.uk}
}

\maketitle
% \date{}

\begin{abstract}
We propose a new method for estimating the extreme quantiles for a function
of several dependent random variables. In contrast to the conventional approach
based on extreme value theory, we do not impose the condition that
the tail of the underlying distribution admits an approximate parametric form,
and, furthermore, our estimation makes use of the full observed data.
The proposed method is semiparametric as no parametric forms
are assumed on all the marginal distributions. But we select appropriate
bivariate copulas to model the joint dependence structure by taking the
advantage of the recent development in constructing large dimensional
vine copulas. Consequently a sample quantile resulted from a large bootstrap
sample drawn from the fitted joint distribution is taken as the estimator
for the extreme quantile. This estimator is proved to be consistent.
% as long as the quantile to be estimated is not too extreme.
The reliable and robust
performance of the proposed method is further illustrated by simulation.

% of several dependent random variables, multivariate extreme value theory
% has been employed by either modeling tail dependence function
% parametrically or estimating tail dependence function nonparametrically
% and then using its homogeneous property to extrapolate data. However,
% when the number of dependent variables is not small, using multivariate
% extreme value theory becomes practically infeasible.  Given recent
% development in constructing large dimensional Vine copulas, we propose a
% semiparametric method to estimating the extreme quantile by modeling
% dependence via copulas and marginals nonparametrically when the quantile
% is not too extreme. Empirical study shows that the proposed method
% performs quite well and robust.
\end{abstract}

\bigskip

\noindent
{\sl Keywords}:
Bootstrap, D-vine copula, empirical distribution function,
extreme quantile, sample quantiles, time series.

\newpage

\section{Introduction}

Let $\{\bX_1, \cdots, \bX_n\}$ be a sample from the population of
a $p$-variate random vector $\bX = (X_1, \cdots, X_p)$.
Let $\xi = h(\bX)$ be a random variable defined as a function of
$\bX$, where the function $h(\cdot)$ is known.
The goal of this paper is to estimate the $(1-\alpha)$-th quantile
of $\xi$, i.e.
\begin{equation} \label{a1}
Q_{\xi}(\alpha) = \min\{\, x: P( \xi \le x ) \ge 1- \alpha \,\},
\end{equation}
where $\alpha > 0$ is a very small constant such that $n\alpha$ is small.
When $\alpha < 1/n$, $Q_{\xi}(\alpha)$ is outside the range of observed data.
This rules out the possibility to estimate $Q_{\xi}(\alpha)$
by the sample quantile of  $\{ \xi_1, \cdots, \xi_n\}$, where
$\xi_i= h(\bX_i)$. This study was motivated by practical problems in financial
risk management. For example, a `traffic light' stress-test
requires to alarm `red light scenario' when, for example, a
test metric $\xi = h(X_1, \cdots, X_p)$ crosses over its $(1-\alpha)$-th quantile with $\alpha
=0.0005$ or 0.0001, while $X_1, \cdots, X_p$ are the prices of a trade along different
tenors (from 3 days to 25 years). The size of available data paths is
typically in the order of a few thousands.

The standard approach to estimate quantiles outside the range of the data
is to assume that the distribution
of $\xi$ is in the domain of attraction of an extreme value distribution.
Based on the characterization of this assumption (Proposition 3.3.2 of Embrechts,
Kl\"uppelberg and Mikosch 1997), extreme quantiles can be estimated via the estimation
for the parameters in the extreme value distribution and the normalized constants.
However the estimation is inefficient as only a small proportion of the
observations at a tail can be used. This  causes further difficulties
in practice as the estimation is often sensitive to the proportion of the data used.
See, e.g., section~6.4.4 of Embrechts,
Kl\"uppelberg and Mikosch (1997) for a detailed account of this approach.

In addition to the methods based on univariate extreme value theory, one
can also assume that $\bX$ lies in the domain of attraction of a multivariate
extreme value distribution; see de Haan and Ferreira (2006). This
implies that the tail distribution of each component of $\bX$
can be approximated by a parametric form determined by an extreme value
distribution while the joint tail dependence has a nice homogeneous property.
For estimating extreme quantiles for the functions of $\bX$, one can
model the joint tail dependence either parametrically (Coles and Tawn 1994)
or nonparametrically, and then extrapolate data based on the homogeneous property
(de Haan and Sinha 1999, and Drees and de Haan 2013). Although using multivariate
extreme value theory may be more efficient than using univariate extreme value
theory (Bruun and Tawn 1988), the sensitivity on the amount data used in estimation
remains as a serious drawback. Furthermore, when the dimension of $\bX$ is not small,
finding a parametric family for the joint tail dependence is extremely difficult
and the nonparametric estimation for the joint tail dependence becomes
too poor to be practically usable.

% In fact it is an almost impossible task when $p=1$ unless some strong
% assumption is imposed on the underlying distribution.
% However the task become
% feasible when $p>1$ as we do not necessarily have to go to extremes at each
% marginal distributions in order to observe a (jointly) extreme event.
% Let's look at the simple example: suppose $X_1,\cdots,X_p$ are a random sample from the
% uniform distribution on $[0, 1]$ and $h(\bX)=\min\{X_1,\cdots,X_p)$. Then
% $P(h(\bX)>1-n^{-\gamma/p})=n^{-\gamma}$ for any $\gamma>0$. Therefore,
% when $\gamma\in (1, p)$, $P(X_i>1-n^{-\gamma/p})=o(n^{-1})$. That is, if
% we know the dependence (independence in this example) and $\alpha$ is not
% small enough ($\gamma<p$ in this example), then we should be able to estimate the quantile without
% fitting a parametric family to marginals. This motivates us to fit a
% parametric family to the copula of $\bX$ and to model marginals
% nonparametrically.

In this paper, we propose a new  semiparametric method for estimating
$Q_{\xi}(\alpha)$.
It consists of three steps: (i) we apply the empirical distribution transformation
to each components of $\bX$ to make all the marginal distributions approximate
$U[0, 1]$, (ii) we then select an appropriate copula
to model the joint dependence structure, (iii) finally we draw a large bootstrap sample
$\{ \bX_1^\star, \cdots, \bX_m^\star\}$
from the fitted joint distribution derived from (i) and (ii),
and estimate $Q_{\xi}(\alpha)$ by the $(1-\alpha)$-th sample quantile of
$\{ \xi_1^\star, \cdots, \xi_m^\star \}$, where $\xi_i^\star=
h(\bX_i^\star)$. Fitting a $p$-dimensional copula in (ii) is feasible due to
the recent development of vine copula construction; see section 2 below.
The bootstrap sample size $m$ can be arbitrarily large. In practice we typically
require, e.g. $m\alpha \ge 20$.
This method does not impose a parametric form directly on the tail of the distribution
of $\xi$ or $\bX$. It makes use of the whole available data, and, hence, provides more
robust performance than the methods based on extreme value theory.

It is a known fact that $Q_\xi(\alpha)$ can be well estimated by the
$(1-\alpha)$-th sample quantile even when $\alpha \to 0$ but
$\alpha n \to \infty$; see Theorem 3.1 of
Dekkers and de Haan (1989). Our method  is somehow in this spirit.
The fact that $Q_\xi(\alpha)$ depends on $p$ variables with $p>1$ makes it possible
to generate a bootstrap sample of size $m$ greater, or much greater, than $n$.
Although our method can handle the cases when the components of $\bX$ are dependent
with each other, its essence is at its clearest when all $X_1, \cdots, X_p$ are independent,
as then a bootstrap sample for $\bX$ can be easily obtained by sampling each component
separately from its $n$ observations. Note that the corresponding bootstrap sample space
consists of $n^p$ elements. It ensures sufficient diversity in the bootstrap sample even
for $m$ much greater than $n$.

However the fundamental reason for our approach to be a creditable one is that it is not
necessary to go to extremes along any component of $\bX$ in order to
observe a joint extreme event. We report a simple simulation result below to illustrate this
key point. Let all components  $X_j$ be i.i.d., and
$\xi = {1\over p} \sum_{1\le j \le p} X_j$.
We approximate the probability
$\alpha = P\{ \xi > Q_\xi (\alpha) \}$ by
\[
\wh\alpha_n = P\{\; \xi > Q_\xi (\alpha), \; F^{-1}_j(1/n) \le X_j \le F^{-1}_j(1-1/n) \;\;
{\rm for} \; 1\le j \le p \; \},
\]
where $F_j(\cdot)$ denotes the marginal distribution function of $X_j$.
With available $n$ observations, the distribution range for
$X_j$ covered by the data can be regarded as from $F^{-1}_j(1/n)$ to
$F^{-1}_j(1-1/n)$.  This range cannot be enlarged by resampling from the
observed data. Thus $\wh\alpha_n$ can be regarded as  the probability of
the event $\{ \xi > Q_\xi (\alpha) \}$
truncated within the range covered by a sample of size $n$.
Our method will work when $\wh\alpha_n$ is close to $\alpha$,
as we can only model the joint distribution
well within the observed range.

The table below lists the values of $\wh\alpha_n$  calculated by a simulation
with 1,000,000 replications for $p=20$, $n=500$ or $1,000$ and the distribution of $X_j$ being
uniform on the unit interval, standard normal or Student's $t$ with 4 degrees freedom.
 Note that $t_4$ is a very
heavy-tailed distribution, as $E(X_j^4) = \infty$ if $X_j \sim t_4$.

\begin{center}
\begin{tabular}{cc|ccccc}
Distribution of $X_j$ &  $n$& $\alpha =.05$ & $\alpha =.01 $& $\alpha
=.005$&$\alpha =.001$&$\alpha =.0005$\\
\hline
$U(0,1)$& 500 &.04741 &.00942 &.00436 &.00078 &.00045\\
        & 1000 &.04809 &.00949 &.00438 &.00084 &.00046\\
\hline
$N(0,1)$&    500& .04360 & .00829 & .00401 & .00075 &.00038\\
& 1000& .04645 &.00896 &.00439 &.00083 &.00043\\
\hline
$t_4$ & 500 & .03629 &.00540 &.00204 &.00013&.00004\\
      & 1000& .04183 &.00609 &.00251 &.00020 &.00005\\
\end{tabular}
\end{center}

This simulation indicates
that it is possible to estimate $Q_\xi (\alpha)$ accurately for $\alpha$ as small
as 0.0005 even with sample size $n=500$ when $\bX$ is
uniformly distributed or normal. However for the heavy-tailed
distributions such as $t_4$, the proposed method may incur large
estimation errors, and therefore is not adequate. In fact our approach does
not involve any direct extrapolations,
it can estimate extreme but {\sl not too extreme} quantiles. How extreme it can go depends
on the underlying distribution, the sample size $n$, and the form of function $h(\cdot)$ which
defines $\xi$.
However when $\xi$ is defined in terms of
empirical marginal distribution functions,
all marginal distributions are effectively $U(0,1)$. Then our method will
provide accurate estimation (see also section 4 below). The multiple comparison methods
 based on marginal $P$-values fall into this category.

The rest of the paper is organized as follows. The methodology is presented in section 2.
It also contains a brief introduction of D-vine copulas. The asymptotic properties are
developed in section 3. Simulation illustration is reported in section 4.

% uniform distribution on $[0, 1]$ and $h(\bX)=\min\{X_1,\cdots,X_p)$. Then
% $P(h(\bX)>1-n^{-\gamma/p})=n^{-\gamma}$ for any $\gamma>0$. Therefore,
% when $\gamma\in (1, p)$, $P(X_i>1-n^{-\gamma/p})=o(n^{-1})$.

\section{Methodology}

\subsection{Notation}

Let $\bX = (X_1, \cdots, X_p)$, $F(\cdot)$ be the cumulative distribution
function (CDF) of $\bX$,
 $F_j(\cdot)$ be the CDF of $X_j$,
and $U_j=F_j(X_j)$. Then $U_j \sim U[0, 1]$ for $1\le j \le p$.
Let $\bX_i= (X_{i1}, \cdots, X_{ip})$, $i=1,\cdots,n$, be a random sample from $\bX$. Put
\begin{align} \label{b1}
\wh F_j(x) = {1 \over {n+1}} \sum_{i=1}^n I(X_{ij} \le x), \qquad
U_{ij} = \wh F_j(X_{ij}).
\end{align}
Then $\sup_x|\wh F_j(x)-F_j(x)|\overset{p}{\to}0$, and
 $\{ U_{1j}, \cdots, U_{nj}\}$ may be
{\sl approximately} regarded as a sample from $U[0,1]$ when $n$ is large.

It follows from Sklar's theorem that for $\bx = (x_1, \cdots, x_p) \in R^p$,
\begin{align} \label{b2}
&
F(\bx) =
P(X_1 \le x_1, \cdots, X_p \le x_p)\\
  =& P\{ U_1 \le F_1(x_1),
\cdots, U_p \le F_p(x_p)\}
= C\{ F_1(x_1), \cdots, F_p(x_p) \},
\nonumber
\end{align}
where $C(\cdot)$ is the CDF of $\bU \equiv (U_1, \cdots, U_p)$, and is called a
$p$-variate copula. In fact $C(\cdot)$ is a distribution function on
 $[0, 1]^p$ with all one-dimensional uniform marginal distributions.
We always assume that $C(\cdot)$ admits a probability density function (PDF),
denoted by $c(\cdot)$, which is called a copula density function.
Then the joint PDF of $\bX$ can be written as
\begin{equation} \label{b2n}
f(\bx) = c\{ F_1(x_1), \cdots, F_p(x_p) \} \prod_{i=1}^p f_j(x_j),
\end{equation}
where $f_j(\cdot)$ is the PDF of $X_j$.
Hence $c(\cdot) \equiv 1$ if and only if $X_1, \cdots, X_p$ are independent.
For more properties on copulas we refer to Nelson (2006).  Due to the invariant property
with respect to marginals, copula models have become one of the most frequently
used tool in risk management; see McNeil, Frey and Embrechts (2005).

\subsection{Estimation for $F(\cdot)$}
Representations (\ref{b2}) and (\ref{b2n}) separate the dependence among the components
of $\bX$ from the marginal distributions. They
 indicate clearly that the dependence is depicted by a copula.
A natural and completely nonparametric estimator for  the copula function $C(\cdot)$
 is the empirical copula function
\begin{equation} \label{b3}
\wh C(\bu) = {1 \over n} \sum_{i=1}^nI( U_{i1} \le u_1, \cdots, U_{ip} \le u_p ), \qquad
\bu = (u_1, \cdots, u_p) \in [0, 1]^p.
\end{equation}
Obviously such a nonparametric estimator $\wh C(\cdot)$ suffers from the so-called
`curse-of-dimensionality' even for moderately large $p$, though it is still
root-$n$ consistent; see, e.g. Fermanian \etal (2004). One alternative is to
impose the assumption that the unknown copula  belongs to a parametric family
$\{ c(\cdot; \, \btheta), \; \btheta \in \bTheta \}$, where copula density
function $ c(\cdot; \, \btheta)$
is known upto the $d$ unknown parameters $\btheta$, the parameter space
$\bTheta$ is a subset of $R^d$ and $d\ge 1$ is an integer.
Then $\btheta$ can be estimated %, at least in principle,
by, for example, the maximal likelihood
estimator defined as
\begin{equation} \label{b4}
\wh \btheta = \arg\max_{\btheta} {1 \over n}
\sum_{i=1}^n \log c(U_{i1}, \cdots, U_{ip}; \btheta).
\end{equation}
See also section 2.3 below for further discussion  on the specification
of $c(\cdot; \btheta)$. Now
by (\ref{b2}), an estimator for the CDF of $\bX$  is defined as
\begin{equation} \label{b5}
\wh F (\bx) = C\{ \wh F_1(x_1), \cdots, \wh F_p(x_p); \wh \btheta \}, \qquad \bx \in R^p,
\end{equation}
where $C(\cdot; \, \btheta)$ is the CDF corresponding to the PDF $c(\cdot; \, \btheta)$.

\subsection{Copula specification: D-vines}

For any integer $p \ge 3$, a $p$-variate copula function can be effectively
specified via pairwise decomposition, leading to various forms of vine
copulas (Bedford and Cooke 2001, 2002). Different orders of the pairings in the
decomposition yield different vines. Nevertheless, only
bivariate copula functions are to be specified.
When the components of random
vector $\bX$ (therefore also $\bU$) are naturally ordered (such as time
series), the D-vine copulas
are particularly easy to use. A copula density function, i.e. a PDF of $\bU$,
% the joint density function of $\bU=(U_1, \cdots, U_p)$
specified by a D-vine admits the form
\begin{equation} \label{b6}
c(\bu)=
\prod_{j=1}^{p-1} \prod_{i=1}^{p-j} c_{i,\, i+j|i+1, \cdots, i+j-1}
\{ F(u_i|u_{i+1}, \cdots, u_{i+j-1}), \; F(u_{i+j}|u_{i+1}, \cdots, u_{i+j-1})\},
\end{equation}
see, for example, (8) of Aas \etal (2009), where $F(u_k|u_{i+1}, \cdots, u_{i+j-1})$
denotes the conditional CDF of $U_k$ given $(U_{i+1}=u_{i+1}, \cdots, U_{i+j-1} =
u_{i+j-1})$, and $c_{i,\, i+j|i+1, \cdots, i+j-1}(\cdot)$ denotes the copula
density for the conditional distribution of $(U_i, U_{i+j})$ given
$U_{i+1}, \cdots, U_{i+j-1}$.
Now some remarks are in order.

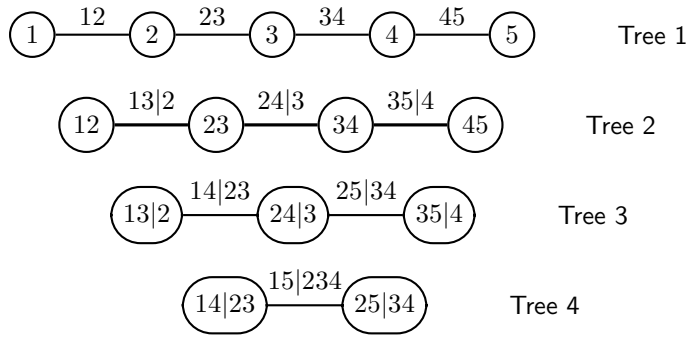
\begin{figure}[htb]
\begin{center}
{\footnotesize
 \begin{tikzpicture}[node distance=3em,scale=0.8,font=\sffamily\small,auto]
\tikzstyle{ibox}=[draw=black,thick,shape=circle,text badly centered];
\tikzstyle{gbox}=[draw=black,thick,shape=rectangle,rounded corners=1.2em,inner sep=4.5pt,
minimum height=2.3em,text badly centered];
\node[ibox] (a1) at (0,0) {$1$};
\node[ibox, right= of a1] (a2) {$2$};
\node[ibox, right= of a2] (a3) {$3$};
\node[ibox, right= of a3] (a4) {$4$};
\node[ibox, right= of a4] (a5) {$5$};
\node[right= of a5]  {Tree 1};
\draw[-, thick] (a1) -- node[above,sloped] {\texttt{$12$}} (a2);
\draw[-, thick] (a2) -- node[above,sloped] {\texttt{$23$}} (a3);
\draw[-,thick] (a3) -- node[above,sloped] {\texttt{$34$}} (a4);
\draw[-,thick] (a4) -- node[above,sloped] {\texttt{$45$}} (a5);
\node[ibox] (a6) at (0.9,-1.5) {$12$};
\node[ibox, right= of a6] (a7) {$23$};
\node[ibox,right= of a7] (a8) {$34$};
\node[ibox,right= of a8] (a9) {$45$};
\node[right= of a9]  {Tree 2};
\draw[-,very thick] (a6) -- node[above,sloped] {\texttt{$13|2$}} (a7);
\draw[-,very thick] (a7) -- node[above,sloped] {\texttt{$24|3$}} (a8);
\draw[-,very thick] (a8) -- node[above,sloped] {\texttt{$35|4$}} (a9);
\node[gbox] (v1) at (1.9,-3)  {$13|2$};
\node[gbox, right= of v1] (v2) {$24|3$};
\node[gbox,right= of v2] (v3) {$35|4$};
\node[right= of v3]  {Tree 3};
\draw[-, thick] (v1) -- node[above,sloped] {\texttt{$14|23$}} (v2);
\draw[-, thick] (v2) -- node[above,sloped] {\texttt{$25|34$}} (v3);
\node[gbox] (v4) at (3.2,-4.5)  {$14|23$};
\node[gbox, right= of v4] (v5) {$25|34$};
\node[right= of v5]  {Tree 4};
\draw[-, thick] (v4) -- node[above,sloped] {\texttt{$15|234$}} (v5);
\end{tikzpicture}
}
\caption{Tree illustration of a D-Vine with 5 variables.}\label{fig:1}
\end{center}
\end{figure}

\askip

\noindent
{\bf Remark 1}.
(i) Only bivariate copula density functions are used in (\ref{b6}).
See Joe (1997) for various parametric copula families which can be used to
specify those copula functions.

(ii)
A $p$-variate D-vine can be represented as a graph with the maximum $p$-1
trees, corresponding to $j=1, \cdots, p-1$ on the RHS of (\ref{b6}); see,
for example, Aas \etal (2009).
However the construction of those trees must be done in
the order of $j=1, 2, \cdots, p-1$. For example,
the conditional CDF $F(u_i|u_{i+1}, \cdots, u_{i+j-1} )$ is required in
the $j$-th tree. By Lemma 1 below, it can be calculated based on a copula
constructed in the ($j$-1)th tree:
\[
F(u_i|u_{i+1}, \cdots, u_{i+j-1} ) =
{ \partial C_{i,i+j-1|i+1, \cdots, i+j-2}\{ F(u_i|u_{i+1}, \cdots, u_{i+j-2}),
\; F(u_{i+j-1}| u_{i+1}, \cdots, u_{i+j-2}) \}
\over
\partial F(u_{i+j-1}| u_{i+1}, \cdots, u_{i+j-2}) },
\]
where $C_{i,i+j-1|i+1, \cdots, i+j-2}(\cdot)$ is the copula corresponding
to the copula density $c_{i,i+j-1|i+1, \cdots, i+j-2}(\cdot)$ specified in
the ($j$-1)th tree. For $j=1$, $F(u_i) = u_i$. For $j=2$,
\[
F(u_i|u_{i+1}) =
{\partial C_{i, i+1} \{ F(u_i), \; F(u_{i+1}) \} \over
\partial F(u_{i+1}) }
= {\partial C_{i, i+1} (u_i, u_{i+1}) \over \partial u_{i+1} }.
\]
Figure~\ref{fig:1} illustrates the tree structure of a D-vine with $p=5$ variables.

(iii) $U_i$ and $U_{i+j}$ are conditionally independent given $U_{i+1}, \cdots,
U_{i+j-1}$ if and only if
\begin{equation} \label{b8}
c_{i,\, i+j|i+1, \cdots, i+j-1}(\cdot) \equiv 1.
\end{equation}
This follows from (\ref{b2n}) by letting $f(\bx)$ be the
conditional PDF of $(U_i, U_{i+j})$ given
$U_{i+1}, \cdots, U_{i+j-1}$.

(iv)
In applications we often assume that the dependence is of the order $m (<p)$ in the sense
that (\ref{b8}) holds for all $j> m$.  Then (\ref{b6}) reduces to
\begin{equation} \label{b9}
c(\bu)=
\prod_{j=1}^{m} \prod_{i=1}^{p-j} c_{i,\, i+j|i+1, \cdots, i+j-1}
\{ F(u_i|u_{i+1}, \cdots, u_{i+j-1}), \; F(u_{i+j}|u_{i+1}, \cdots, u_{i+j-1})\}.
\end{equation}
A particular simple case is a Markov D-vine copula which admits the dependence
at order $m=1$ with the copula density function of the form
\begin{equation} \label{b10}
c(\bu)=  \prod_{i=1}^{p-1} c_{i,i+1}\{ F(u_i), F(u_{i+1})\}=
 \prod_{i=1}^{p-1} c_{i,i+1} (u_i, u_{i+1}),
\end{equation}
where $c_{i,j} (\cdot)$ are bivariate copulas. For example, when the components of
$\bX$ are $p$ successive values of a Markov process,
% AR(1) time series (with independent innovations),
$\bX$ admits a Markov D-vine copula.

(v)
We may apply some goodness-of-fit statistics to choose among different
specifications or to test a particular model. The goodness-of-fit can be
measured in terms of the difference between the empirical copula
$\wh C(\cdot)$ defined in (\ref{b3}) and
the fitted parametric copula $C(\cdot ;\btheta)$ in (\ref{b5}).
This leads to the Kolmogorov-Smirnov and Cram{\'e}r-von Mises statistics
\begin{equation} \label{b11}
T_{n} = n \int_{[0,1]^p} \big\{ C(\bu; \wh \btheta) - \wh C(\bu) \big\}^2
d \bu, \qquad
S_n = \sup_{\bu \in [0,1]^p} \sqrt{n} \big|C(\bu; \wh \btheta) - \wh C(\bu)\big|.
\end{equation}
Genest and R{\'e}millard  (2008) showed that both the above
statistics lead to a consistent test in the sense that
if the true copula is not within the specified parametric family,
the model will be rejected with probability converging to 1.
Unfortunately their asymptotic null distributions depend on
the underlying distribution. In practice the parametric bootstrap
method described in Appendix A of Genest \etal (2009) can be used
to evaluate the $P$-values. The validity of the bootstrap method is established by
Genest and R{\'e}millard  (2008).

(vi) The D-vine decomposition (\ref{b6}) is valid for any continuous distribution
on $[0,1]^p$ with uniform marginal distributions.  On the other hand, with
any bivariate copula density functions used on the RHS of (\ref{b6}), the
D-vine constructed in the manner described in (ii) above is a valid $p$-variate copula,
i.e. (\ref{b6})  is a proper PDF on $U[0,1]^p$ with uniform marginals.
Both these assertions can be established
by mathematical induction.
% Proof for the 1st statement:
% It is trivially true for
% $p=2$. Suppose it is also true for $p=k$,
% we now show it for $p=k+1$.
% Let $\bu_k = (u_1, \cdots, u_k)$ and $\bu_{k+1} = (\bu_k, u_{k+1})$.
% By the Bayesian formula, it holds that
% \[ c(\bu_{k+1} ) = c(\bu_k) f(u_{k+1} |\bu_k). \]
% Since (\ref{b6}) holds for $c(\bu_k)$, it will also hold for $c(\bu_{k+1} )$ if we
% can prove that
% \begin{align} \label{b11n}
% &
% f(u_{k+1} |\bu_k)  = f(u_{k+1}|u_1, \cdots, u_k)\\
%  =& \prod_{j=1}^k c_{k+1-j,\, k+1|k+2-j, \cdots, k}
% \{ F(u_{k+1-j}|u_{k+2-j}, \cdots, u_k), \; F(u_{k+1}|u_{k+2-j}, \cdots, u_k)\}.
% \nonumber
% \end{align}
% Applying (\ref{b2n}) with $f(\bx)$ being the conditional PDF of $(U_1, U_{k+1})$
% given $(U_2, \cdots, U_k)$ yields
% \begin{align*}
% &
% f(u_{k+1}|u_1, \cdots, u_k)\\  =& f(u_{k+1}|u_2, \cdots, u_k)
% \cdot c_{1,\, k+1|2, \cdots, k} \{ F(u_1|u_2, \cdots, u_k), \;
% F(u_{k+1} |u_2, \cdots, u_k) \}.
% \end{align*}
% Now (\ref{b11n}) follows from applying above decomposition iteratively $k$ times.
%
% For the 2nd, we only need to show that the expression for $c(\bu_{k+1})/c(\bu_k)
% is a conditional PDF of $U_{k+1}|\bU_k$.

(vii) When the components of $\bX$ are not naturally orders as a time series, other
vine copula families such as C-vine could be used. We refer to
Czado,  Brechmann and Gruber (2013) for a survey on the selection of vine copulas.

\askip

\noindent
{\bf Lemma 1}. Let $Y$ and $Z$ be two random variables, $\bW$ be a random
vector, and $\bZ = (Z, \bW)$. Denoted by, respectively, $F_\bW$ and $C_\bW$
the CDF and the copula of $\bW$. Then it holds that
\begin{equation} \label{b12}
F_{Y|Z} (y|z) = {\partial C_{Y,Z}\{ F_Y(y), F_Z(z) \}
\over
\partial F_Z(z)}, \quad
F_{Y|\bZ} (y|\bz) = {\partial C_{Y,Z|\bW}\{ F_{Y|\bW} (y|\bw), F_{Z|\bW} (z|\bw) \}
\over
\partial F_{Z|\bW}(z|\bw)}.
\end{equation}

\askip

First equality in (\ref{b12}) follows from calculus. The second equality
follows from the first by applying it to the conditional distribution of
$(Y, Z)$ given $\bW$. Those relationships were first established by Joe~(1996).

\subsection{Estimation for extreme quantiles} % for functions of $\bX$}

With the estimated distribution (\ref{b5}) for $\bX$, in principle we can
deduce an estimator for the distribution of $\xi = h(\bX)$. Unfortunately
in most applications such an estimator cannot be evaluated explicitly.
We propose to draw a bootstrap sample $\bX_1^\star, \cdots, \bX_m^\star$ from (\ref{b5}),
and to estimate the extreme quantile $Q_\xi(\alpha)$ of $\xi$ (see (\ref{a1}))
by the corresponding sample quantile of $\{ \xi_i ^\star = h(\bX_i^\star) \} $, i.e.
\begin{equation} \label{b14}
\wh Q_\xi(\alpha) = \xi^\star_{[m\alpha]},
\end{equation}
where $\xi^\star_{[j]}$ denotes the $j$-th largest value among $ \xi_1 ^\star,
\cdots, \xi_m ^\star$.
We require $m$ sufficiently large such that, for example, $m\alpha \ge 20$.

We apply the inverse of the Rosenblatt transformation to draw
$u_1, \cdots, u_p$ from
D-vine copula density (\ref{b6}).
Then we let
\begin{equation} \label{b15}
x_j = \wh F_j^{-1} (u_j), \quad j =1, \cdots, p,
\end{equation}
where $\wh F_j$
defined in (\ref{b1}). To this end, draw $v_1, \cdots, v_p$ independently
from $U[0,1]$. Let
$u_1 =v_1$, and
\[
u_i = F^{-1}(v_i|u_1, \cdots, u_{i-1}) \qquad {\rm for} \; i=2, \cdots, p,
\]
where $F^{-1}(\cdot \, | u_1, \cdots, u_{i-1})$ denotes the inverse function
of the conditional CDF of $U_i$ given $(U_1=u_1, \cdots, U_{i-1}=u_{i-1})$
which is determined by the D-vine copula density (\ref{b6}). It follows from
Lemma 1 that
\[
F(u_i|u_1, \cdots, u_{i-1})
= { \partial C_{1,\, i|2,\cdots, i-1}\{ F(u_1|u_2, \cdots, u_{i-1}), \;
F(u_i|u_2, \cdots, u_{i-1}) \} \over
\partial F(u_1|u_2, \cdots, u_{i-1}) },
\]
where $C_{1,\, i|2,\cdots, i-1}(\cdot)$ is the copula function corresponding to
the copula density $c_{1,\, i|2,\cdots, i-1}$ contained on the RHS of (\ref{b6}).
Aas \etal (2009) outlined an algorithm to implement the above scheme.

\askip

\noindent
{\bf Remark 2}. When all the components of $\bX$ are known to be
independent with each other, our approach still applies. In this case,
$\bX^\star_i = (X_{i1}^\star, \cdots, X_{ip}^\star)$ can be obtained
with $X_{ij}^\star$ resampled independently from $\{ X_{1j}, \cdots, X_{nj}\}$.

\section{Asymptotic properties}

In this section we present the consistency for our extreme quantile estimation.
Recall $C(\cdot) = C(\cdot; \btheta)$ is the CDF of $\bU = (U_1,\cdots,U_p)$.
The target quantile, as a function of $\btheta$, can be expressed as
\[
Q_{\xi}(\alpha;\btheta)= \min\big\{ x: P_{\btheta}( \xi > x) \le  \alpha \big\},
\]
where $\xi = h(\bX)= h\{F_1^{-1}(U_1),\cdots,F_p^{-1}(U_p)\}$; see (\ref{a1}).
Put
\begin{align*}
& A(x)=\big\{(u_1,\cdots,u_p): h\{F_1^{-1}(u_1),\cdots,F_p^{-1}(u_p)\}>x\big\},\\
& A_n(x)=\big\{(u_1,\cdots,u_p): (u_1,\cdots,u_p)\in A(x),\; \frac1{n+1}\le
u_1, \cdots, u_p \le \frac{n}{n+1}\big\},\\
& B_n(x)=\Big\{(u_1,\cdots,u_p): h\{F_1^{-1}(\hat
G_1^{-1}(u_1)),\cdots,F_p^{-1}(\hat G_p^{-1}(u_p))\}>x,\\
& \quad \hspace{3.8cm} \frac 1{n+1}\le
u_1,\cdots,u_p\le \frac{n}{n+1}\Big\},
\end{align*}
where $\hat
G_{j}(x)=\frac 1{n+1}\sum_{i=1}^nI(U_{ij}\le x)$, and $U_{ij}$ is defined in
(\ref{b1}).
Let $\btheta_0$ denote the true value of $\btheta$.
Hence $Q_{\xi}(\alpha)=Q_{\xi}(\alpha;\btheta_0) $ is
the true quantile to be estimated.
As we estimate extreme quantiles, we assume $\alpha \equiv \alpha_n \to 0$ as
$n \to \infty$.

Some regularity conditions are now in order.

\begin{itemize}
\item [A1.] $||\hat\btheta-\btheta_0||=O_p(\Delta_n)$ for some $\Delta_n\to 0$ as $n\to\infty$.
\item [A2.] For any constant $M>0$,  if
\[\sup_{||\btheta-\btheta_0||\le M\Delta_n}\Big|\alpha_n^{-1} \int_{A(x_n(\btheta))}c(u_1,\cdots,u_p;\btheta)\,du_1\cdots
du_p-1\Big|\to 0\] and
\[\sup_{||\btheta-\btheta_0||\le M\Delta_n} \Big|\alpha_n^{-1}
\int_{A(y_n(\btheta))}c(u_1,\cdots,u_p;\btheta)\,
du_1\cdots du_p-1 \Big|\to 0\] for sequences $x_n(\btheta)$ and
$y_n(\btheta)$ as $n\to\infty$,  then
$\sup_{\btheta\in\bTheta}|x_n(\btheta)/y_n(\btheta)-1|\to 0$ as $n\to\infty$.

\item [A3.] For any constant $M>0$, if
\[\sup_{||\btheta-\btheta_0||\le M\Delta_n}\Big|\alpha^{-1}_n
\int_{B_n(x_n(\btheta))}c(u_1,\cdots,u_p;\btheta)\,du_1\cdots du_p-1
\Big|\overset{p}{\to} 0\]
and
\[\sup_{||\btheta-\btheta_0||\le M\Delta_n}\Big| \alpha_n^{-1}
\int_{B_n(y_n(\btheta))}c(u_1,\cdots,u_p;\btheta)\,du_1\cdots du_p-1
\Big|\overset{p}{\to} 0\] for sequences $x_n(\btheta)$ and $y_n(\btheta)$
as $n\to\infty$,  then
$\sup_{\btheta\in\bTheta}|x_n(\btheta)/y_n(\btheta)-1|\overset{p}{\to} 0$ as $n\to\infty$.

\item [A4.] As $n \to \infty$, it holds for any constant $M>0$ that
 $$\sup_{||\btheta-\btheta_0||\le M\Delta_n} \Big|\frac{\int_{B_n(Q_{\xi}(\alpha;\btheta))}c(u_1,\cdots,u_p;\btheta)\,
du_1\cdots du_p}{\int_{A_n(Q_{\xi}(\alpha;\btheta))}c(u_1,\cdots,u_p;\btheta)\,
du_1\cdots du_p}-1 \Big|\overset{p}{\to} 0.$$

\item [A5.] As $n \to \infty$, it holds for any constant $M>0$ that
$$\sup_{||\btheta-\btheta_0||\le M\Delta_n} \Big|
\frac{\int_{A_n(Q_{\xi}(\alpha;\btheta))}c(u_1,\cdots,u_p;\btheta)\,
du_1\cdots du_p}{\int_{A(Q_{\xi}(\alpha;\btheta))}c(u_1,\cdots,u_p;\btheta)\,
du_1\cdots du_p}-1 \Big|\to 0.$$

\item [A6.]  As $n \to \infty$, it holds  for any constant $M>0$ that
$$\sup_{||\btheta-\btheta_0||\le M\Delta_n}
\Big| \alpha_n^{-1} \int_{A(Q_{\xi}(\alpha))}c(u_1,\cdots,u_p;\btheta)\,
du_1\cdots du_p-1 \Big|\to 0.$$
\end{itemize}

\askip

\noindent{\bf Theorem 1.} Under Conditions A1--A6, $\hat
Q_{\xi}(\alpha)/Q_{\xi}(\alpha)\overset{p}{\to} 1$ as $n\to\infty$.

\askip

\noindent{\bf Proof.}  Note that
\begin{equation}
\label{pfTh-1}\alpha=\int_{A(Q_{\xi}(\alpha;\btheta))}
c(u_1,\cdots,u_p;\btheta)\,du_1\cdots du_p\end{equation}
and $\hat Q_{\xi}(\alpha)$ satisfies
\begin{equation}
\label{pfTh-2}\int_{B_n(\hat
Q_{\xi}(\alpha))}c(u_1,\cdots,u_p;\hat\btheta)\,du_1\cdots
du_p/\alpha=1+o_p(1).\end{equation}

Write
\[\begin{array}{lll}
&\int_{B_n(\hat
Q_{\xi}(\alpha))}c(u_1,\cdots,u_p;\hat\btheta)\,du_1\cdots du_p-\alpha\\
=&\int_{B_n(\hat
Q_{\xi}(\alpha))}c(u_1,\cdots,u_p;\hat\btheta)\,du_1\cdots
du_p-\int_{B_n(Q_{\xi}(\alpha;\hat\btheta))}c(u_1,\cdots,u_p;\hat\btheta)\,
du_1\cdots du_p\\
&+\int_{B_n(Q_{\xi}(\alpha;\hat\btheta))}c(u_1,\cdots,u_p;\hat\btheta)\,
du_1\cdots du_p-\int_{A_n(Q_{\xi}(\alpha;\hat\btheta))}c(u_1,\cdots,u_p;\hat\btheta)\,
du_1\cdots du_p\\
&+\int_{A_n(Q_{\xi}(\alpha;\hat\btheta))}c(u_1,\cdots,u_p;\hat\btheta)\,
du_1\cdots du_p-\int_{A(Q_{\xi}(\alpha;\hat\btheta))}c(u_1,\cdots,u_p;\hat\btheta)\,
du_1\cdots du_p.
\end{array}\]
Then it follows from (\ref{pfTh-1}), (\ref{pfTh-2}) and Conditions A1, A4, A5 that
\begin{align}
\label{pfTh-3}
&\frac{1}{\alpha}\int_{B_n(\hat Q_{\xi}(\alpha))}c(u_1,\cdots,u_p;\hat\btheta)\,
du_1\cdots du_p\overset{p}{\to} 1, \quad\text{and}\quad\\
&
\frac{1}{\alpha}\int_{B_n( Q_{\xi}(\alpha;\hat\theta))}c(u_1,\cdots,u_p;\hat\btheta)\
du_1\cdots du_p\overset{p}{\to}1
\nonumber
\end{align}
as $n\to\infty$. By (\ref{pfTh-3}) and Condition A3, we have
\begin{equation}
\label{pfTh-4}
\hat Q_{\xi}(\alpha)/Q_{\xi}(\alpha;\hat\theta)\overset{p}{\to} 1
\end{equation}
as $n\to\infty$.  It follows from (\ref{pfTh-1}), Conditions A1, A2 and A6  that
\begin{equation}
\label{pfTh-5}
Q_{\xi}(\alpha;\hat\theta)/Q_{\xi}(\alpha)\overset{p}{\to} 1.
\end{equation}
Hence, the theorem follows from (\ref{pfTh-4}) and (\ref{pfTh-5}).
\qed

\askip

\noindent{\bf Remark 3.}
Condition A1 holds with $\Delta_n=1/\sqrt n$ under some regularity conditions as in Genest, Ghoudi
and Rivest (1995).
Condition A2 implies that the extreme quantile is asymptotically uniquely determined.
Condition A3 implies that  the extreme quantile is still asymptotically
uniquely determined when the marginal distributions are replaced by their empirical
counterparts. Condition A4 ensures that sets $A_n$ and $B_n$ are close enough. Condition A5
ensures that there is no need to extrapolate the marginal distributions
below $\hat G_i^{-1}(\frac 1{n+1})$ and above $\hat G_i^{-1}(\frac n{n+1})$.
We illustrate those conditions in two examples below.

\askip

\noindent{\bf Example 1: Gumbel Copula.} Suppose the distribution of
$\bX$ is the Gumbel copula
\[C(x_1,\cdots,x_p;\theta)=\exp\big\{-\big(\sum_{i=1}^p(-\log
x_i)^{\theta}\big)^{1/\theta}\big\},\]
where $\theta>0$. Consider $h(\bX)=\{\max_{1\le i\le p}X_i\}^{-1}$
 and
$\alpha=n^{-\gamma}$ for some $\gamma>1$. Then
$Q_{\xi}(\alpha;\theta)=n^{\gamma/p^{1/\theta}}$ and
$Q_{\xi}(\alpha)=Q_{\xi}(\alpha;\theta_0)$. It is easy to check that for any $i=1,\cdots,p$
\[P\big\{X_i\le n^{-1}, X_j\le Q_{\xi}^{-1}(\alpha;\theta)\quad\text{for}\quad j=1,\cdots,i-1,i+1,\cdots, p\}=n^{-(1+\gamma^{\theta}(p-1)/p)^{1/\theta}}.\] So
when $\gamma<p^{1/\theta}$, we have
\[P(X_i\le n^{-1}, X_j\le Q_{\xi}^{-1}(\alpha;\theta)\quad\text{for}\quad  j=1,\cdots,i-1,i+1,\cdots, p)/\alpha\to 0,\]
 which can be used to prove Condition A5. It is straightforward to
verify Conditions A1, A2 and A6  when $\gamma\in (1,  \,p^{1/\theta})$. Use
the fact that
\begin{equation}
\label{ex0}
\sup_{u}\Big|\frac{\sqrt n(\hat
G_i^-(u)-u)}{u^{\delta}(1-u)^{\delta}}I\big(\frac 1{n+1}\le u\le\frac
n{n+1}\big)\Big|=O_p(1)
\end{equation} for any $\delta\in (0, 1/2)$, we can show that for any
$\epsilon\in (0, 1)$, the following relation
\[A_n\{(1-\epsilon)x_n(\theta)\}\subset B_n(x_n(\theta))\subset
A_n\{(1+\epsilon)x_n(\theta)\}\]
holds with probability tending to one for any sequence
$x_n(\theta)/Q_{\xi}(\alpha;\theta)$ converging to a positive constant.
By the above relation, one can show Conditions A3 and A4 hold when
$\gamma\in (1, \, p^{1/\theta}).$

\askip

\noindent{\bf Example 2: Clayton copula.} Suppose the distribution of
$\bX$ is \[F(x_1,\cdots,x_d;\theta,\beta)=(1-p+\sum_{i=1}^px_i^{-\beta\theta})^{-1/\theta}\]
for some $\theta>0$ and $\beta>0$. % is given.
Then the copula of $\bX$ is the Clayton copula
\[C(u_1,\cdots,u_p;\theta)=(1-p+\sum_{i=1}^pu_i^{-\theta})^{-1/\theta}.\]
 Consider $h(\bX)=\{\max_{1\le i\le p}X_i \}^{-1}$ and
$\alpha=n^{-\gamma}$ for some $\gamma>1$. Then
$Q_{\xi}(\alpha;\theta)=(\frac{n^{\gamma \theta}-1+p}{p})^{1/(\beta\theta)}$
and $Q_{\xi}(\alpha)=Q_{\xi}(\alpha;\theta_0)$.
It is easy to check that for any $i=1,\cdots,p$
\begin{equation}\label{Ex2-1}\begin{array}{ll}
&P(X_i\le n^{-1},
X_j\le Q_{\xi}^{-1}(\alpha;\theta)\quad\text{for}\quad j=1,\cdots,i-1, i+1,\cdots,p)\\
=&\{1-p+n^{\beta\theta}+
\frac{(p-1)(n^{\theta\gamma}-1+p)}p\}^{-1/\theta}.\end{array}\end{equation}
When $\gamma<\beta$, the right hand side of (\ref{Ex2-1}) is $o(n^{-\gamma})$, which can be used to show Condition A5 holds.
The rest conditions can be verified as Example 1 when $1<\gamma<\beta$.
When the distribution of $\bX$ is Clayton copula, i.e.,
$\beta=1$ for the above distribution, the right hand side of (\ref{Ex2-1}) is the same order as $n^{-\gamma}$, which implies that Condition A5 does not hold. That is, the marginals have to be modeled parametrically for estimating this extreme quantile with $\alpha=n^{-\gamma}$  in this case.

\askip

Theorem 1 above is generic, imposing the conditions directly
on the closeness of between the quantile set $A$ and
its truncated version $A_n$, the empirical approximation $B_n$ for $A_n$.
When the copula of $\bX$ is multivariate regular variation (i.e. Condition B2 below)
and the quantile set $A$ is scalar-invariant (see Condition B1 below),
Theorem 2 below shows that the consistency still holds.

\begin{itemize}
% \item [Bi)] $||\hat\btheta-\btheta_0||=O_p(\Delta_n)$ for some $\Delta_n\to 0$.
\item [B1.]
Let $S\subset (0, 1]^p$ be a set independent of $n$, $\beta>0$ be a constant,
 and $0\le a_n\to 0$ be any such
a sequence.
 When
% $Q_{\xi}(\alpha;\btheta)\to a<\infty$ as $n\to\infty$ (i.e.,
% the right endpoint of the CDF of $\xi$ is finite),
$Q_{\xi}(0;\btheta) = a<\infty$,
put $\bar a_n(\btheta)=a-Q_{\xi}(\alpha;\btheta)$ and assume
$A(a-a_n)=a_n^{\beta}S$.
When $Q_{\xi}(0;\btheta)=\infty$,
% $Q_{\xi}(\alpha;\btheta)\to\infty$ as $n\to\infty$,
put $\bar a_n(\theta)=1/Q_{\xi}(\alpha;\theta)$ and assume
% for any $a_n\to0$, $A(a_n^{-1})=a_n^{\beta}S$ for some $\beta>0$ and set
% $S\subset (0, 1]^p$ independent of $n$.
$A(a_n^{-1})=a_n^{\beta}S$.

\item [B2.] For any $M>0$, there exists $N$ such that, as $t\to0$
\[\sup_{n\ge N}\sup_{||\btheta-\btheta_0||\le
M\Delta_n}
\Big|\frac{c(tu_1,\cdots,tu_p;\btheta)}{c(t,\cdots,t;\btheta)}-
l(u_1,\cdots,u_p;\btheta) \Big|\to 0\]  for $u_1,\cdots,u_p>0$, and
\[\sup_{n\ge N}\sup_{||\btheta-\btheta_0||\le
\delta_0} \Big|\frac{c(tu,\cdots,tu;\btheta)}{c(t,\cdots,t;\btheta)}-u^{\gamma} \Big|=0\]
for $u>0$ and some $\gamma\in R$. Further
$$\sup_{n\ge
N}\sup_{||\btheta-\btheta_0||\le
M\Delta_n}\int_Sl(u_1,\cdots,u_p;\btheta)\,du_1\cdots du_p<\infty.
$$

\item [B3.] For any $M>0$,  \[\sup_{||\btheta-\btheta_0||\le
M\Delta_n} \Big|\frac{c(\bar a_n^{\beta}(\btheta_0),\cdots, \bar
a_n^{\beta}(\btheta_0);\btheta)}{c(\bar
a_n^{\beta}(\btheta_0),\cdots,\bar
a_n^{\beta}(\btheta_0);\btheta_0)}-1 \Big|\to0\] as $n\to\infty$.

\item [B4.] $\lim_{n\to\infty}\sup_{||\btheta-\btheta_0||\le
M\Delta_n}(n^{\delta}\bar a_n^{\beta}(\btheta))\in (0, \infty)$ for some
$\delta\in (0, 1)$.
\end{itemize}

\askip

\noindent{\bf Theorem 2.} Under Conditions A1 and B1--B4,  $\hat
Q_{\xi}(\alpha)/Q_{\xi}(\alpha)\overset{p}{\to} 1$ as $n\to\infty$.

\askip

\noindent{\bf Proof.} We shall verify conditions A2--A6 in Theorem 1. By B1, we can write
\[\begin{array}{ll}
\alpha&=\int_{A(Q_{\xi}(\alpha;\btheta))}c(u_1,\cdots,u_p;\btheta)\,du_1\cdots du_p\\
&=\int_{\bar a_n^{\beta}(\btheta)S}c(u_1,\cdots,u_p;\btheta)\,du_1\cdots du_p\\
&=\int_Sc(\bar a_n^{\beta}(\btheta)u_1,\cdots,\bar a_n^{\beta}(\btheta)u_p;\btheta)\bar a_n^{\beta p}(\btheta)\,du_1\cdots du_p
\end{array}\]
and then it follows from A1 and B2 that
\begin{equation}
\label{pfTh2-1}
\frac{\alpha}{c(\bar a_n^{\beta}(\btheta),\cdots,\bar a_n^{\beta}(\btheta);\btheta)\bar a_n^{\beta p}(\btheta)}=\int_S l(u_1,\cdots,u_p;\btheta)\,du_1\cdots du_p.
\end{equation}
Like the proof of (\ref{pfTh2-1}), condition A2 can be shown by using B2.
Note that B1 and B4 imply that $A_n(Q_{\xi}(\alpha;\btheta))=A(Q_{\xi}(\alpha;\btheta))$ for $||\btheta-\btheta_0||\le M\Delta_n$ and large $n$. Hence Condition A5 holds. Using (\ref{ex0}) we can show condition A4. Note that
$\alpha^{-1}\int_{B_n(x_n(\btheta))}c(u_1,\cdots,u_p;\btheta)\to 1$ implies that $x_n(\btheta)\to Q_{\xi}(\alpha;\btheta)$. Hence, like the proof of (\ref{pfTh2-1}), we can show A3 by using (\ref{ex0}), B1 and B2.
Condition A6 follows from B2 and B3. Hence, Theorem 2 follows from Theorem 1.
\qed

\askip

\noindent{\bf  Remark 4.} Condition B1 relates the set A to a fixed set
$S$ by a scaling factor depending on the sample size $n$. This idea
appeared in Drees and de Haan (2013). Condition B2 assumes the copula
density is a multivariate variation. We refer to Resnick (1987) for more
details on multivariate regular variation. It follows from Condition B2 that
$c(\bar a_n^{\beta}(\btheta_0),\cdots,\bar
a_n^{\beta}(\btheta_0);\btheta)=O(\bar
a_n^{\beta\gamma-\epsilon}(\btheta_0))$ for any $\epsilon>0$. Hence,
(\ref{pfTh2-1}) implies $$\alpha= \alpha_n = O(\bar
a_n^{\beta(\gamma+p-\epsilon)}(\btheta_0))$$ for any $\epsilon>0$. This
reflects the fact that how small $\alpha_n$ can be depends on
the geometry of the set A ($\beta$), the
property of the copula ($\gamma$) and the dimension ($p$). It is
straightforward to check that Conditions B1--B4 hold for the above two examples on
Gumbel copula and Clayton copula with $\beta>\gamma$ for $\alpha=n^{-\gamma}$.

\section{Numerical properties}

In this section we illustrate the proposed method by simulation.
We let $\bX = (X_{1}, \cdots, X_{p})'$, where
\begin{equation} \label{d1}
X_{t} = 1.2 X_{t-1} - 0.6 X_{t-2} + \ve_{t},
\end{equation}
and $\ve_{t}$ are independent and identically distributed random variables.
We estimate the extreme quantiles of the following four functions:
\begin{align*}
h_1(\bX) & = X_{(p)} + X_{(p-1)} + X_{(p-2)}, \qquad
 h_2(\bX)  = \min_{1\le t \le p} F_t(X_t), \\
h_3 (\bX) &= {1 \over p} \sum_{t=1}^p X_t, \qquad \qquad
\qquad \qquad\,
 h_4(\bX)  = {1 \over p} \sum_{t=1}^p \{1 - F_t(X_t)  \} ,
\end{align*}
where $X_{(1)} \le \cdots \le X_{(p)}$ are the order statistics of the components of
$\bX$, $F_t(\cdot)$ is the CDF of the $t$-th component of $\bX$, and hence
$F_t(X_t) \sim U(0,1)$.

We consider two distributions for $\ve_t$ in (\ref{d1}), namely the standard normal $N(0,1)$,
and Student's $t$-distribution with 4 degrees of freedom $t_4$.
% , and log standard normal.
With a sample $\bX_1, \cdots, \bX_n$ drawn from the distribution of $\bX$, we estimate
the $(1-\alpha)$-th quantile with  $\alpha =0.05$, 0.01, 0.005, 0.001 and 0.0005.
We set the sample size $n=500$ or 1,000, and the dimension $p=20$ or 40.
For each sample, we fit the data with three D-vine copulas:
\begin{quote}
Copula I: \;\;\;  two trees only (i.e. $m=2$ in (\ref{b9})) with Gaussian
binary copulas.

Copula II: \;\;  two trees only with all binary copulas selected by the AIC.

Copula III: \; the number of trees and all binary copulas are selected by the AIC.
\end{quote}
Since $X_t \sim \AR(2)$ (see (\ref{d1})), $X_t$ and $X_{t+3}$ are independent conditionally
on $X_{t+1}$ and $X_{t+2}$. Hence the dependence structure of $\bX$ can be represented by
a D-vine with two trees, i.e. Copula II reflects the underlying dependence structure correctly.
Furthermore Copula I specifies the correct parametric model when $\ve_t
\sim N(0, 1)$ in (\ref{d1}).

The computation was carried out using the R-package {\tt CDVine} which
selected binary copulas from a large number of copula families; see {\tt
cran.r-project.org/web/packages/CDVine/CDVine.pdf}. We let $m=40,000$ in (\ref{b14}).

For each setting, we drew 400 samples, i.e. replicated the estimation 400 times.
We calculate the Mean Absolute Relative Error (MARE):
\begin{equation} \label{d2}
\MARE = {1 \over 400} \sum_{i=1}^{400} \Big| { \wh Q_i - Q \over Q} \Big|,
\end{equation}
where $Q$ denotes the true quantile value, and $\wh Q_1, \cdots, \wh Q_{400}$ denote
its estimated values over 400 replications.  The true values of the
extreme quantiles for $h_1(\bX), \cdots, h_4(\bX)$ were calculated by a
simulation with a sample of size 500,000.
For the comparison purpose, we also include the simple sample quantile
estimate $\xi_{[n\alpha]}$ from an original samples, where $\xi_{[j]}$
denotes the $j$-th largest value among
$\xi_k \equiv h_i(\bX_k)$ for $k=1, \cdots, n$, and $i=1,\cdots,4$.

\begin{singlespace}
\begin{table}[tbh]
\begin{center}
\caption{MARE for estimating the $(1-\alpha)$-th quantiles of $h_i(\bX)$ ($i=1, \cdots, 4$) with
 $n=500$, $p=20$ and $\ve_t \sim N(0, 1)$.}
\askip
\label{Tab1}
\begin{tabular}{|c|l|c|c|c|c|c|}
\hline
Function
& Model & $\alpha = .05$ & $\alpha = .01$& $\alpha = .005$ & $\alpha =
.001$ & $\alpha = .0005$\\
\hline
            & Copula I & .0161 & .0259& .0337& .0587& .0721\\
$h_1(\bX)$ & Copula II & .0167 & .0256& .0327& .0603& .0720\\
	       & Copula III& .0169 & .0258 & .0327 & .0597 & .0709\\
           & sample quantile& .0231 & .0373 & .0476 & .0841 & n/a \\
\hline
           & Copula I & .0082 & .0103& .0119& .0151 & .0199\\
$h_2(\bX)$ & Copula II & .0128 & .0125 &.0126& .0169 & .0189\\
           &Copula III& .0138 & .0132 & .0130& .0168& .0213\\
           & sample quantile& .0404 & .0586 & .0718 & .1069 &  n/a\\
\hline
           &Copula I & .0260 & .0216 & .0204 & .0215& .0227\\
$h_3(\bX)$ &Copula II& .0277 & .0253 & .0258 & .0287& .0291\\
           &Copula III& .0289& .0257& .0262 & .0283 & .0293\\
           & sample quantile& .0463 & .0572 & .0632 & .1020 &  n/a\\
\hline
           &Copula I& .0028 & .0035& .0041& .0050& .0064\\
$h_4(\bX)$  &Copula II&.0035 & .0045& .0050& .0056& .0063\\
           &Copula III&.0042&.0051 & .0057& .0066& .0074\\
           & sample quantile& .0097 & .0167 & .0196 & .0328 & n/a \\
\hline
\end{tabular}
\end{center}
\end{table}
\end{singlespace}

\begin{figure}[htb]
\centering
\includegraphics[width=0.9\textwidth]{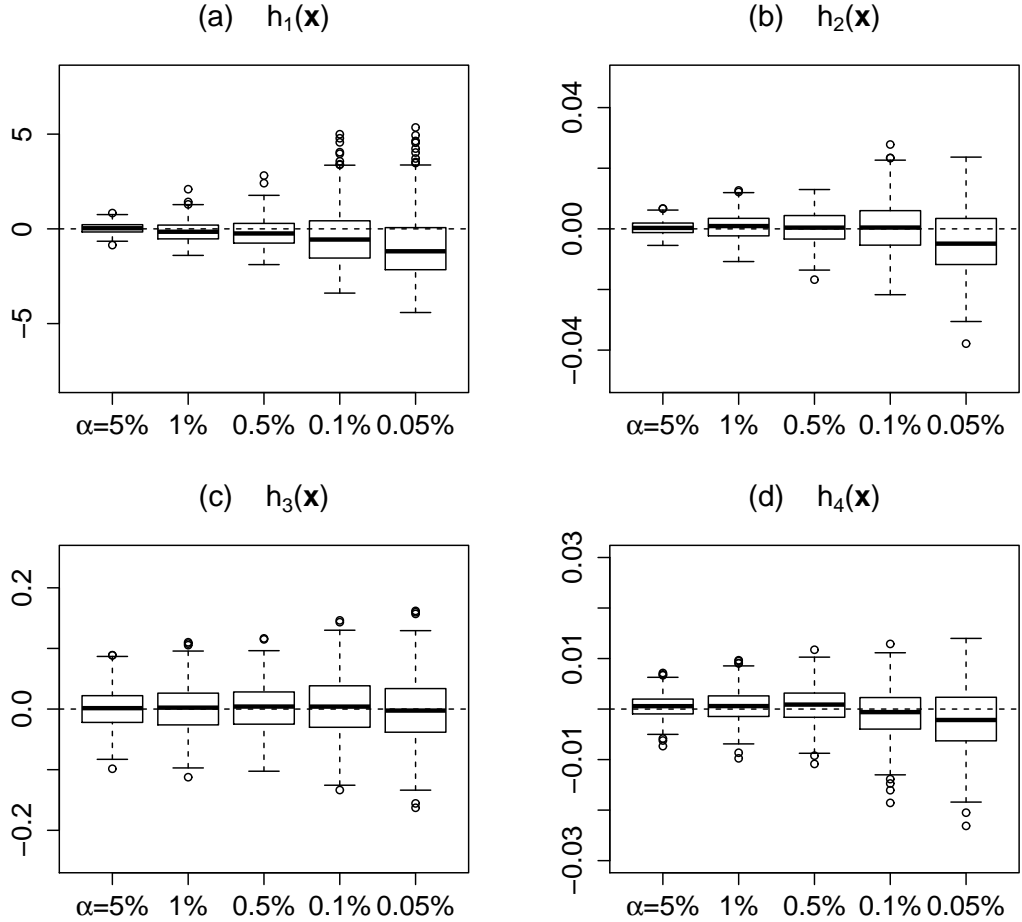}
\caption{Boxplots of the errors in estimating the
$(1-\alpha)$-th quantiles of $h_i(\bX)$ ($i=1, \cdots, 4$) with
$\ve_t \sim N(0, 1)$.
Copula I was used in estimation with $n=500$ and $p=20$. }
\label{Fig2}
\end{figure}

Table~\ref{Tab1} lists the MARE with sample size $n=500$ and $\bX$ consisting of
$p=20$ successive values of the AR(2) process defined by (\ref{d1}) with
standard normal innovations. Since Copula I is the true parametric family for
the underlying distribution, it yields the better estimates than Copulas II and III.
Note that both Copulas II and III are still correct models with more parameters
to be specified. The differences from using three copulas are not substantial;
indicating that the AIC worked well in choosing binary copula
functions (for Copulas II and III) as well as specifying the number of trees (for Copula III).
Also the MARE tends to increase when $\alpha$ decreases; indicating the increasing
difficulty in estimating more extreme quantiles.
In fact we reported in the table the MARE which is defined as the mean
absolute error (MAE) divided by the true quantile value; see (\ref{d2}).
In fact the MAE strictly increases when $\alpha$ decreases.
Figure \ref{Fig2} displays the boxplots of the estimation errors (i.e. $\wh Q_i - Q$,
$i=1, \cdots, 400$; see (\ref{d2})) for the estimation with Copula I, $n=500$ and $ p=20$.
It shows clearly that both the bias and variance of the estimators increase when
$\alpha $ decreases.
Note that $n\alpha$ ranges from 25 to 0.25 for $0.05 \le \alpha \le
0.0005$. For the most extreme case with $\alpha =0.0005$,
we extrapolate far out of the range covered by data $\{ h_i(\bX_t), \; t=1, \cdots, n\}$.
Still the maximum MARE is under 8\% with function $h_1(\bX)$, is under 3\% with
$h_3(\bX)$, and is even smaller with $h_2(\bX)$ and $h_4(\bX)$. We also
notice that the extreme
quantiles of $h_2(\bX)$ and $h_4(\bX) $ can be estimated  much more accurately
than those of $h_1(\bX)$ and $h_3(\bX) $. This is due to the fact that $h_2$ and $h_4$
are the function of the marginal distribution functions of $\bX$. Therefore they are
effectively the functions of a $p$ random vector with all the marginal distributions
being $U(0, 1)$.
% which are always between 0 and 1. Therefore the margins for errors are smaller.
Furthermore,
their estimates do not suffer from the errors due to the inverse empirical
transformations (\ref{b15}) in the bootstrap resampling.
Overall with normal $\bX$, the proposed estimation method works very well.
It provides much more accurate estimates than the simple sample quantiles even
for $\alpha =0.05$ when there are $n\alpha=25$ data points in the top $\alpha$-tails.
% The improvement over the sample quantile estimation increases when $\alpha$ decreases.
With sample size $n=500$ (or even 1000),  the sample quantiles at the $(1-\alpha)$-th level
when $\alpha =0.0005$ are not available.

\begin{singlespace}
\begin{table}[tbh]
\begin{center}
\caption{MARE for estimating the $(1-\alpha)$-th quantiles  of $h_i(\bX)$ ($i=1, \cdots, 4$) with
 $n=500$, $p=20$ and $\ve_t \sim t_4$.}
\askip
\label{Tab2}
\begin{tabular}{|c|l|c|c|c|c|c|}
\hline
Function
& Model & $\alpha = .05$ & $\alpha = .01$& $\alpha = .005$ & $\alpha =
.001$ & $\alpha = .0005$\\
\hline
            & Copula  I & .0277 & .0601& .0918& .2070& .2766\\
$h_1(\bX)$ & Copula  II & .0277 & .0582& .0853& .1887& .2383\\
	       & Copula  III& .0277 & .0576 & .0808 & .1866 & .2399\\
         & sample quantile & .0332 & .0703 & .0950 & .2456 & n/a \\
\hline
           & Copula  I & .0264 & .0114& .0112& .0197 & .0260\\
$h_2(\bX)$ & Copula II & .0094 & .0181 &.0224& .0210 & .0214\\
           & Copula III& .0104 & .0188 & .0227& .0227& .0220\\
           & sample quantile & .0401 & .0579 & .0657 & .1046 & n/a \\
\hline
           & Copula I & .0309 & .0289 & .0334 & .0626& .0848\\
$h_3(\bX)$ & Copula II& .0357 & .0702 & .0897 & .1339& .1405\\
           & Copula III& .0370& .0696& .0904 & .1343 & .1377\\
           & sample quantile & .0496 & .0651 & .0738 & .1569 & n/a \\
\hline
           & Copula I& .0063 & .0080& .0075& .0078& .0089\\
$h_4(\bX)$  & Copula II&.0029 & .0045& .0048& .0067& .0090\\
           & Copula III&.0037&.0051 & .0054& .0075& .0093\\
           & sample quantile & .0089 & .0163 & .0192 & .0305 & n/a \\
\hline
\end{tabular}
\end{center}
\end{table}
\end{singlespace}

\begin{singlespace}
\begin{table}[tbh]
\begin{center}
\caption{MARE for estimating the $(1-\alpha)$-th quantiles of $h_i(\bX)$ ($i=1, \cdots, 4$) with
 $n=500$, $p=40$ and $\ve_t \sim t_4$.}
\askip
\label{Tab3}
\begin{tabular}{|c|l|c|c|c|c|c|}
\hline
Function
& Model & $\alpha = .05$ & $\alpha = .01$& $\alpha = .005$ & $\alpha =
.001$ & $\alpha = .0005$\\
\hline
            &Copula  I & .0290 & .0802& .1234& .2349& .2868\\
$h_1(\bX)$ &Copula  II & .0287 & .0635& .0943& .2095& .2435\\
	       &Copula  III& .0288 & .0635 & .0932 & .2096 & .2424\\
          & sample quantile & .0326 & .0748 & .1014 & .2746 & n/a \\
\hline
           &Copula  I & .0613 & .0379& .0299& .0235 & .0248\\
$h_2(\bX)$ &Copula II & .0255 & .0144 &.0132& .0156 & .0202\\
           &Copula III& .0280 & .0173 & .0157& .0169& .0208\\
           & sample quantile & .0427 & .0654 & .0769 & .1188 & n/a \\
\hline
           &Copula I & .0283 & .0248 & .0269 & .0440& .0604\\
$h_3(\bX)$ &Copula II& .0379 & .0667 & .0861 & .1114& .1179\\
           &Copula III& .0377& .0659& .0864 & .1147 & .1204\\
           & sample quantile & .0486 & .0657 & .0742 & .1350 & n/a \\
\hline
           &Copula I& .0051 & .0069& .0075& .0075& .0066\\
$h_4(\bX)$  &Copula II&.0018 & .0029& .0036& .0045& .0050\\
           &Copula III&.0026&.0041 & .0046& .0053& .0058\\
           & sample quantile & .0065 & .0114 & .0144 & .0253 & n/a \\
\hline
\end{tabular}
\end{center}
\end{table}
\end{singlespace}

\begin{singlespace}
\begin{table}[tbh]
\begin{center}
\caption{MARE for estimating the $(1-\alpha)$-th quantiles of $h_i(\bX)$ ($i=1, \cdots, 4$) with
 $n=1000$, $p=20$ and $\ve_t \sim t_4$.}
\askip
\label{Tab4}
\begin{tabular}{|c|l|c|c|c|c|c|}
\hline
Function
& Model & $\alpha = .05$ & $\alpha = .01$& $\alpha = .005$ & $\alpha =
.001$ & $\alpha = .0005$\\
\hline
            &Copula I & .0225 & .0475& .0689& .1786& .2441\\
$h_1(\bX)$ &Copula II & .0210 & .0422& .0586& .1419& .2043\\
	       &Copula III& .0207 & .0424 & .0587 & .1370 & .2039\\
           & sample quantile& .0256 & .0516 & .0721 & .1630 & n/a \\
\hline
           &Copula I & .0282 & .0111& .0097& .0186 & .0265\\
$h_2(\bX)$ &Copula II & .0079 & .0144 &.0185& .0167 & .0168\\
           &Copula III& .0087 & .0147 & .0186& .0183& .0201\\
           & sample quantile& .0279 & .0407 & .0499 & .0764 & n/a \\
\hline
           &Copula I & .0226 & .0210 & .0258 & .0556& .0808\\
$h_3(\bX)$ &Copula II& .0223 & .0410 & .0563 & .0990& .1128\\
           &Copula III& .0237& .0412& .0577 & .0988 & .1150\\
           & sample quantile& .0351 & .0466 & .0589 & .1065 & n/a \\
\hline
           &Copula I& .0052 & .0069& .0061& .0062& .0077\\
$h_4(\bX)$  &Copula II&.0021 & .0028& .0032& .0050& .0067\\
           &Copula III&.0029&.0037 & .0042& .0055& .0073\\
           & sample quantile& .0061 & .0102 & .0144 & .0258 & n/a \\
\hline
\end{tabular}
\end{center}
\end{table}
\end{singlespace}

Tables \ref{Tab2}--\ref{Tab5} list the MARE when $\ve_t \sim t_4$ in (\ref{d1}). Now components
of  $\bX$ are heavy-tailed with $E(||\bX||^4) = \infty$.  The extreme quantiles
to be estimated are more likely to be impacted by the extreme values of the components
of $\bX$ than the cases with $\ve_t \sim
N(0, 1)$. The MARE with $\alpha=0.001$ and 0.0005 in Tables \ref{Tab2}--\ref{Tab5} tend to
be too large with functions $h_1(\bX)$ and $h_3(\bX)$, while the estimation for those
extreme quantiles of $h_2(\bX)$ and $h_4(\bX)$ remains accurate with the MARE smaller than
3\%. Nevertheless when the sample size increases from $n=500$ to $n=1000$, the MARE decreases.
When the number of components of $\bX$ increases from $p=20$ to $p=40$, the MARE with
$h_1(\bX)$ or $h_2(\bX)$ increases while that with $h_3(\bX)$ and $h_4(\bX)$ decreases.
Note that $h_1(\bX)$ or $h_2(\bX)$ are extreme functions  of the components $\bX$, and
they become more extreme when $p$ increases. In contrast, $h_3(\bX)$ or
$h_4(\bX)$ are the means of the components of $\bX$, they behave more like normal when
$p$ increases due the CLT. With $\ve_t \sim t_4$, Copula I misspecifies
the model while Copula II provides a correct dependence structure (i.e. a
D-vine with two trees only).  With the functions $h_1(\bX), h_2(\bX)$ and $h_4(\bX)$,
the Gaussian copula (i.e. Copula I) is the least preferable,
the estimation with Copula II leads to smaller MARE than those with Copula III across
Tables \ref{Tab2}--\ref{Tab5} although the differences are not substantial, and are certainly
smaller than the differences between the estimates based on Copula II and those based on Copula I.
However with $h_3(\bX)$, the estimation with the Gaussian copula
is the best. One possible explanation is that with $p=20 $ or $p=40$,
it holds approximately that
\[
h_3(\bX) = {1 \over p} \sum_{t=1}^p X_t \;
\sim \; N \big(0,\;  {1\over p} \var(X_1) + {2 \over p} \sum_{k=2}^p{(1-\frac{k-1}{p})}\cov(X_1, X_k) \big).
\]
Since the Gaussian copula
also specifies the correlation among the components of $\bX$ correctly,
it is an approximately correct parametric model. Overall the proposed method
provides more, or much more, accurate estimates than the sample quantiles across
Tables \ref{Tab2}--\ref{Tab5}.

\begin{singlespace}
\begin{table}[tbh]
\begin{center}
\caption{MARE for estimating the $(1-\alpha)$-th quantiles of $h_i(\bX)$ ($i=1, \cdots, 4$) with
 $n=1000$, $p=40$ and $\ve_t \sim t_4$.}
\askip
\label{Tab5}
\begin{tabular}{|c|l|c|c|c|c|c|}
\hline
Function
& Model & $\alpha = .05$ & $\alpha = .01$& $\alpha = .005$ & $\alpha =
.001$ & $\alpha = .0005$\\
\hline
            &Copula  I & .0191 & .0595& .0957& .2167& .2595\\
$h_1(\bX)$ &Copula  II & .0187 & .0426& .0631& .1638& .2292\\
	       &Copula  III& .0181 & .0414 & .0617 & .1673 & .2318\\
          & sample quantile& .0242 & .0557 & .0754 & .1717 & n/a \\
\hline
           &Copula  I & .0618 & .0388& .0298& .0235 & .0242\\
$h_2(\bX)$ &Copula II & .0250 & .0152 &.0136& .0148 & .0180\\
           &Copula III& .0262 & .0158 & .0143& .0163& .0212\\
           & sample quantile& .0309& .0438 & .0559 & .0871 &  n/a\\
\hline
           &Copula I & .0207 & .0195 & .0209 & .0397& .0573\\
$h_3(\bX)$ &Copula II& .0218 & .0412 & .0576 & .0832& .0948\\
           &Copula III& .0226& .0408& .0575 & .0827 & .0990\\
           & sample quantile& .0356 & .0471 & .0542 & .0969 & n/a \\
\hline
           &Copula I& .0047 & .0066& .0070& .0073& .0063\\
$h_4(\bX)$  &Copula II&.0012 & .0020& .0024& .0034& .0042\\
           &Copula III&.0020&.0028 & .0032& .0043& .0052\\
           & sample quantile& .0048 & .0082 & .0106 & .0194 & n/a \\
\hline
\end{tabular}
\end{center}
\end{table}
\end{singlespace}

\section*{Conclusions}

We propose in this paper a new method for estimating the extreme quantiles of
a function of several random variables. The extreme quantiles concerned are
typically outside the range of the observed data. The new method does not
rely on extreme value theory on which the traditional methods are based.
Hence it is more robust and efficient as it utilizes all
the available data and it does not impose any explicit
parametric forms on the tails of the underlying distributions.

The underpinning idea of the new method is that it is not necessary to go to
extremes along any component variable in order to observe a joint extreme event.
This also indicates that the method may be unable to handle the
excessively extreme cases. How extreme
it can do depends on the underlying distribution and the number of the variables
involved. Nevertheless if the function concerned depends on each random variable
through its CDF transformation (such as $h_2(\cdot)$ and $h_4(\cdot)$ used in
section 4), we effectively deal with the cases when all random variables are
bounded. Then the new method can provide accurate estimation for very extreme quantiles.

It is perhaps also worth mentioning a finding from our simulation study.
For the functions in the form $\xi = h(p^{-1} \sum_j g(X_j) )$ with $p$ moderately large,
fitting a Gaussian copula to capture the dependence (i.e. the
correlation) among $g(X_1), \cdots, g(X_p)$ may leads to a satisfactory
estimation for the quantiles of $\xi $. This is due to the fact that
$p^{-1} \sum_j g(X_j)$ would behave like a normal random variable, the
fitted Gaussian copula should provide adequate estimates for its first two moments.

\section*{References}
\begin{singlespace}
\begin{description}
\item
Aas, K., Czado, C., Frigessi, A. and Bakken, H. (2009). Pair-copula constructions of
multiple dependence. {\sl Insurance Mathematics and Economics}, {\bf 44},
182-198.
\item
Bedford, Y. and Cooke, R.M. (2001). Probability density decomposition for
conditional dependent random variables modeled by vinew. {\sl Annals of Mathematics
and Artificial Intelligence}, {\bf 32}, 245-268.
\item
Bedford, Y. and Cooke, R.M. (2002). Vines -- a new graphical model for dependent random
variables. \AS, {\bf 30}, 1031-1068.
\item
 Brechmann, E.C. and Schepsmeier, U. (2013). Modeling Dependence with C- and D-Vine Copulas:
 The R Package CDVine. {\sl  Journal of Statistical Software}, {\bf 52}, 1-27.
\item
Bruun, J.T. and Tawn, J.A. (1998). Comparison of approaches for estimating the probability of coastal flooding. {\sl Appl. Statist.}, {\bf 47}, 405-423.
\item
Coles, S.G. and Tawn, J.A. (1994). Statistical methods for multivariate
extremes: an application to structure design. {\sl J. Royal Statist. Soc. C}, {\bf 43}, 1-31.
\item
Czado, C., Brechmann, E.C. and Gruber, L. (2013).
Selection of vine copulas. In {\sl Copulae in Mathematical and Quantitative Finance}.
Edited by Jaworski, P., Durante, F. and W.K. H\"ardle.
Springer Lecture Notes in Statistics {\bf 213},
pp 17-37.
\item
De Haan, L. and Ferreira, A. (2006). Extreme Value Theory: An Introduction. {\sl Springer.}
\item
De Haan, L. and Sinha, A.K. (1999). Estimating the probability of a rare event. {\sl Ann. Statist.}, {\bf 27}, 732-759.
\item
Dekkers, A.L.M. and de Haan, L. (1989). On the estimation of the extreme-value index and
large quantile estimation \AS, {\bf 17}, 1759-1832.
\item
Drees, H. and de Haan, L. (2013). Estimating failure probabilities. {\sl Bernoulli}, to appear.
\item
Embrechts, P., Kl\"uppelberg, C. and Mikosch, T. (1997). {\sl Modelling Extremal Events}.
Spriner, Berlin.
\item
Fermanian, J.-D., Radulovic, D. and Wegkamp, M. (2004). Weak convergence
of empirical copula processes. {\sl Bernoulli}, {\bf 10}, 847-860.
\item
Genest, C. and Favre, A.-C. (2007). Everything you always wanted to know
about copula modeling
 but were afraid to ask. {\sl Journal of Hydrologic Engineering}, {\bf 12}, 347-368.
\item
Genest, C., Ghoudi K. and Rivest, L.-P. (1995). A semiparametric
estimation procedure of dependence parameters in multivariate families of
distributions. {\sl Biometrika}, {\bf 82}, 543--552.
\item
Genest, C. and R{\'e}millard, B. (2008). Validity of the parametric
bootstrap for goodness-of-fit testing in semiparametric models.
{\sl Ann. Inst. H. Poincar{\'e} Probab. Statist.}, {\bf 44},
1096-1127.
\item
Genest, C., R{\'e}millard, B. and Beaudoin, D. (2009).
Goodness-of-fit tests for copulas: a review and a power study.
{\sl Insurance Mathematics and Economics}, {\bf 44},
199-213.
\item
Joe, H. (1996). Families of $m$-variate distributions with given margins
and $m(m-1)/2$ bivariate dependence parameters. In {\sl Distributions
with Fixed Marginals and Related Topics}. Edited by R\"uschendorf, L., Schweizer, B.
and Taylor, M.D. IMS Lecture Notes {\bf 28}, 120-141.
\item
Joe, H. (1997). {\sl Multivariate Models and Dependence Concepts}.
Chapman \& Hall, London.
\item
McNeil, A.J., Frey, R. and Embrechts, P. (2005). {\sl Quantitative Risk Management: Concepts, Techniques, and Tools.} Princeton University Press.
\item
Nelson, R.B. (2006). {\sl An Introduction to Copulas.} Springer.
\item
Resnick, S.I. (1987). {\sl Extreme Values, Regular Variation, and Point Processes.} Springer.
\item
Rosenblatt, M. (1952). Remarks on a multivariate transformation.
{\sl The Annals of Mathematical Statistics}, {\bf 23}, 470-472.
\end{description}
\end{singlespace}

\bigskip

\noindent
{\bf \large Acknowledgements}. Jinguo Gong was partially supported by
National Social Science Foundation of China (Grant No.12CTJ007).
Qiwei Yao was partially supported by an EPSRC research grant.

\end{document}

%% file: QYalias.tex
 \@addtoreset{equation}{section}
 
\renewcommand{\hat}{\widehat}
\def\singlespace{\def\baselinestretch{1}\@normalsize}

\def\wh{\widehat}

\def\askip{\vspace{0.1in}}

\newcommand{\AR}{{\rm AR}}

\newcommand{\cov}{{\rm Cov}}

\newcommand{\etal}{\mbox{\sl et al.\;}}

\newcommand{\var}{\mbox{Var}}

\newcommand{\ve}{{\varepsilon}}

\newcommand{\bU}{{\mathbf U}}

\newcommand{\bW}{{\mathbf W}}
\newcommand{\bX}{{\mathbf X}}

\newcommand{\bZ}{{\mathbf Z}}

\newcommand{\bu}{{\mathbf u}}

\newcommand{\bw}{{\mathbf w}}
\newcommand{\bx}{{\mathbf x}}

\newcommand{\bz}{{\mathbf z}}

\newcommand{\bTheta} {\boldsymbol{\Theta}}

\newcommand{\btheta} {\boldsymbol{\theta}}

\newcommand{\qed}{\hfill $\blacksquare$ \vspace{3mm}}

\def\6bullets{\bullet\bullet\bullet\bullet\bullet\bullet}

\def\AS{{\sl The Annals of Statistics}}